\documentclass[sigconf]{acmart}
\pdfoutput=1
\renewcommand\footnotetextcopyrightpermission[1]{} 
\usepackage{balance}
\usepackage{booktabs} 

\usepackage{graphicx}
\usepackage{amsmath}
\usepackage{algorithm}
\usepackage[noend]{algpseudocode}

\def\last {{\em last.fm}~}
\def\echo {{\em EchoNest}}

\begin{document}
\title{Temporal Proximity induces Attributes Similarity}
\author{Arun Kumar}
\affiliation{%
  \institution{Dept. of Computer Science}
  \streetaddress{University of Minnesota Twin Cities}
}
\email{kumar250@umn.edu}

\author{Karan Aggarwal}
\affiliation{%
  \institution{Dept. of Computer Science}
  \streetaddress{University of Minnesota Twin Cities}
}
\email{aggar081@umn.edu}

\author{Paul Schrater}
\affiliation{%
  \institution{Depts. of Computer Science and Psychology}
  \streetaddress{University of Minnesota Twin Cities}
}
\email{schrater@umn.edu}

\begin{abstract}

Users consume their favorite content in temporal proximity of consumption bundles according to their preferences and tastes. Thus, the underlying attributes of items implicitly match user preferences, however, current recommender systems largely ignore this fundamental driver in identifying matching items. In this work, we introduce a novel temporal proximity filtering method to enable items-matching. First, we demonstrate that proximity preferences exist. Second, we present an induced similarity metric in temporal proximity driven by user tastes and third, we show that this induced similarity can be used to learn items pairwise similarity in attribute space. The proposed model does not rely on any knowledge outside users' consumption bundles and provide a novel way to devise user preferences and tastes driven novel items recommender.

\end{abstract}

\maketitle

\section{Introduction}
\label{sec:introduction}

Music industry is in a phase of massive shift in the listening styles of the music seekers, corroborated 
by the cheaper hosting availability and shift towards mobile devices like smart phones. With these
fundamental technologies in place to bridge the gap between content providers and seekers, 
past decade has seen huge shift towards digital music subscription platforms like \textit{Pandora} and \textit{Spotify} as well as crowd-sourcing platforms like \textit{8tracks} providing direct access to the consumers to works of independent artists or bands. 

With such a fundamental shift in the market as well as the consumer behaviours in the ``Big" Data Generation, the recommender systems have not been able to address some fundamental issues that worked well in the era
of a selected popular artists. The traditional collaborative filtering 
\cite{sarwar2001item} recommender systems recommend users' based on users' with a similar history of consumption.
 Content based recommender systems \cite{lops2011content}, target the
musical attributes or genres, recommending songs similar to what user's listening profile history would suggest in the attribute space. Hybrid recommender systems utilize both these approaches in varying degrees
to recommend songs. 

These methods suffer from the ``long tail" effect~\cite{Celma:Springer2010} where the recommendor end up recommending popular songs or artists mainly, hence limiting the options of users, which can lead to attrition of users having faced boredom with the platform. However, these systems miss a fundamental fact---users' tend to listen songs 
together---songs complement each other. Also, users' listen songs in themes, depending on the context they 
reside in, like work place, study, and mood, not mutually exclusive. These themes have been inferred by 
providers to be represented in the genre space for a long time. There has been a push for crowd sourced
currations of themed songs, leading to platforms like \textit{8tracks} providing under-represented tunes~\cite{economist} boasting of a user base of 5 million. 

\begin {figure}[h!]
\centering
\includegraphics[width=0.9\linewidth,height=5.05cm]{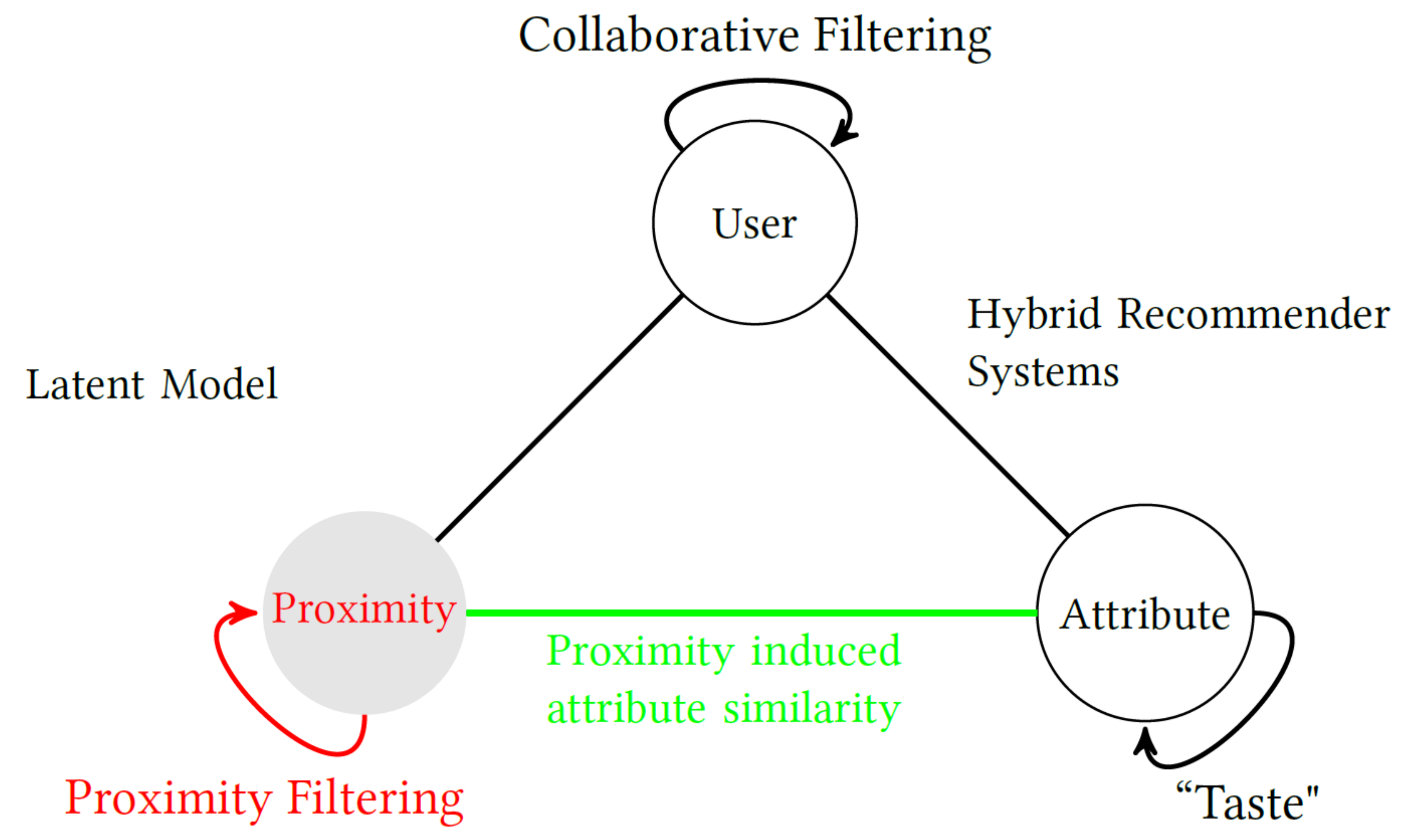}
\caption{Collaborating filtering methods rely on finding other users with common items consumption choices to suggest items interesting to a given user. Content based methods from delve into attributes to find new items with similar attributes as previously consumed by a user. Hybrid systems combine these two methods. We present a novel proximity filtering method to learn attribute similarities.}
\label{fig:overview}
\end{figure}
In this work, we seek to create a fundamental 
framework for a system that can bridge the gap of content based liking as well as capturing the user context using latent themes, bereft of hard genre categorization. As illustrated in figure \ref{fig:overview}, we introduce a key concept of proximity filtering and attempt to answer questions like `what is a better way to find out which items a user might like?' Current systems try to address this question by utilizing collaborative filtering, learning about other users who might have a similar interest and leverage on their items basket to recommend new items. Other collaborative filtering variants leverage additional information with user attributes like age, location etc or explicit feedback for items. For newly released items, consumption patterns  might not be readily available.

However, a close observation of users' music listening pattern, people listen to their favourite songs in \textit{temporal proximity}, provides us insights to learn from their preferences and it can drive identification of new recommendable items. A user's affinity to particular item attributes like beats or rhythms in musical compositions can reveal more into their latent thematic tastes. It resonates with an understanding that tastes or liking are more abstract compositional constructs. We consider an inherently deeper relationship between users' higher level latent thematic tastes and temporal proximity in their consumed contents. These consumption bundles could be key drivers for new generation recommender systems by exploiting this immense amount of unobservable and indirect knowledge of users' liking to induce similarity in attributes and find matching items.

We aim to fill this fundamental gap by introducing a novel paradigm called `Temporal Proximity Filtering' by providing a way to learn similarity in items directly from their attributes, intrinsically driven from their consumption patterns than exterior information. The remainder  of  this  article  details the model. Section \ref{sec:background} provides background. Section \ref{sec:tempproximity} explains the main proximity filtering concepts. Section \ref{sec:attrsimilarity} describes induced attribute similarity. Experimental procedure on music listening dataset is detailed in Section \ref{sec:experiment}. Finally, we discuss the results in section \ref{sec:discussion} and conclude in section \ref{sec:conclusion}.

\section{Background}
\label{sec:background}
One of the main goals of recommender systems is to suggest interesting \emph{novel} items to its users. There are primarily two paradigms: collaborative filtering methods and content based methods. A majority of systems rely on collaborative filtering techniques \cite{sarwar2001item}, following the availability of user provided ratings and community data. Collaborative methods identify users whose item consumption or choices history is similar to a given user and filter recommendable contents from similarly consumed items sets. Such methods largely assume that a given user might also like items from this set because their past choices have commonalities. On the other hand, content based methods~\cite{lops2011content} attempt to find items similar in content to previously liked item by a user. Here, emphasis is on examining items' history for a given user itself and find new items that might be interesting to the user. In music domain, collaborating filtering methods are widely used for providing recommendations, although, content based systems found their usage in a number of applications \cite{lops2011content}. There have been attempts to combine both collaborating and content methods into hybrid methods \cite{melville2002content} so that item contents can augment collaboratively ranked items. Although powerful, a majority of them ignore that user preferences are dynamic and abstract. The methods that can learn from users' content consumption in temporal proximity and abstract level could be beneficial in finding likable items. We attempt to address this aspect by learning attributes similarities and these similarities are hypothesized to be induced by temporal proximity in a latent thematic space.

In domains like music, volume of available contents is growing rapidly, it is often difficult for people to manually select their preferred contents. Although there is a big boost in availability but selection difficulties encourage improvements in preferences driven systems. Contents with diversity and small durations of each items like music have intrinsically different consumption patterns \cite{schedl2015}. Moreover, consumption of such items is more unpredictable compared to items like books etc as there can be long gaps between listening to a similar item again or one can listen to the same music item many times within a shorter time span. Therefore, understanding consumption in a latent thematic space similar to \cite{kumar2017novelty} becomes more important for domains with huge diversity.  

Luke et al. \cite{Barrington08smarterthan} evaluate different music recommendation approaches demonstrating that collaborative filtering based recommender systems produce better recommendations than the ones based on purely artist similarity or acoustic content similarity. They also imply that similarities between song contents can be captured and attempt to find musical cues capturing music similarity augmenting traditional collaborating filtering techniques. Our work emphasizes that user's consumption bundles in a latent thematic space can provide a meaningful way to find items that match user preferences. 

Overload of information led to development of information filtering and retrieval methods that can supply sets of novel items for recommendation. User preferences elicitation plays a key role in identifying such items. Hanani et al.\cite{hanani2001information} suggested usage of explicit elicitation like ratings or implicit elicitation from user behaviors. Meta data for user profiles \cite{Barrington08smarterthan} generally helps in collaborating filtering approaches. Jawaheer et al. \cite{Jawaheer2010} demonstrates that explicit and implicit feedbacks can work in a complementary fashion and they present techniques to leverage both. Bogdanov et al. \cite{bogdanov2013semantic} infer user preferences from explicitly available information. However, these works ignore that preferences are dynamic and abstract. In this paper, we present methods to learn content similarity which are based on users' dynamic taste bundles.

Collaborating filtering methods work really well when system has gathered item information like feedback from other users. With ever evolving online media domains like music, new items are being released constantly. A user might like some of these newly released items but absence of user feedback creates difficulties for collaborating filtering methods. On the other hand, content based methods are capable of recommending items that do not have any prior explicit feedback like rating or comments. However, it is often difficult to craft feature sets and methods to derive identification of recommendable items and requires domain knowledge. Basic premise of these methods lie on information retrieval to filter smaller subset of items from a large set. Moreover, a different set of attributes could be dominant in different items, for example, music might have varying compositions making such retrieval process non-trivial. 

Therefore, information retrieval to acquire knowledge about concerned items requires incorporation of learning techniques. Given a teaching signal, a learning system learns underlying attributes representation and is helpful in predicting a candidate pool of items, potentially novel and interesting to users. This work presents novel similarities metric and learning method to retrieve recommendable items.

\section{Temporal Proximity Filtering}
\label{sec:tempproximity}
\begin{figure}[h!]
\centering
\includegraphics[width=1.0\linewidth,height=3.75cm]{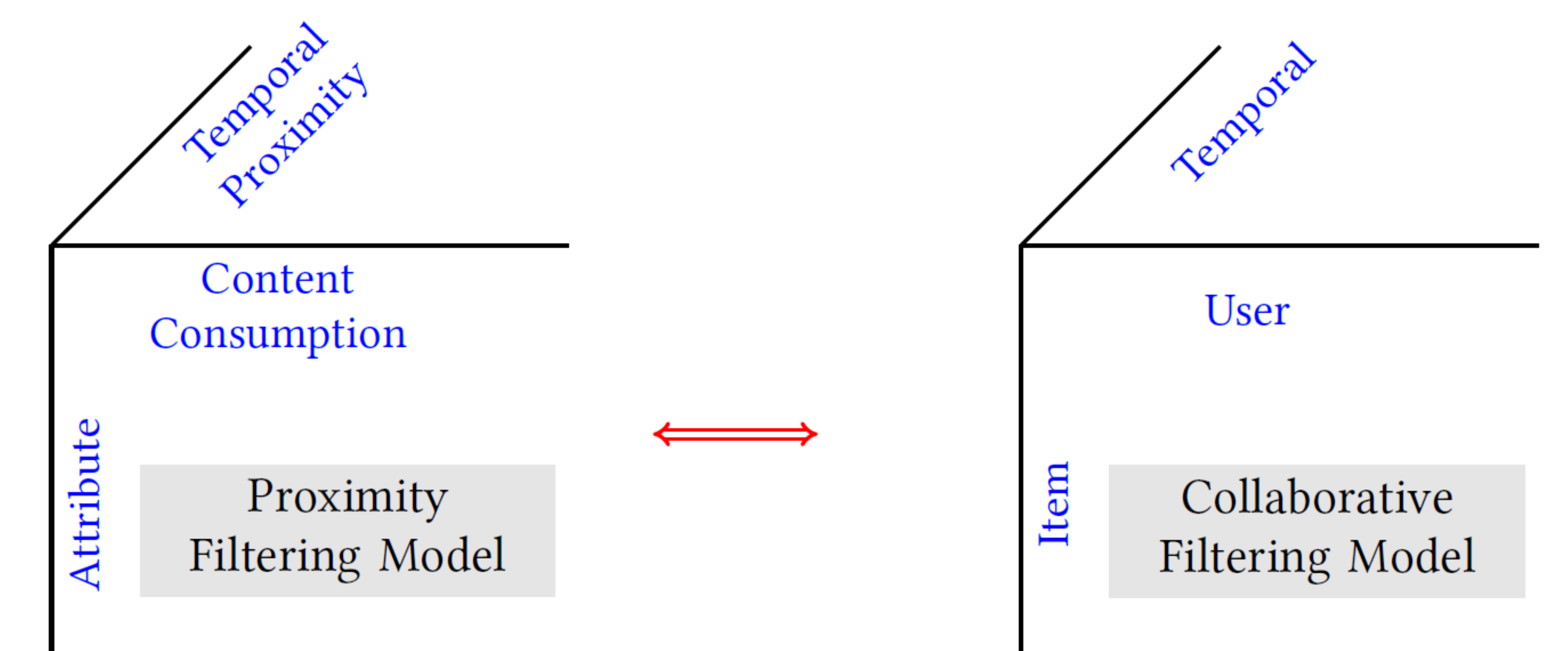}
\caption{Analogy of proposed proximity filtering method with collaborating filtering: Collaborating filtering methods rely on similarity in user profiles and items frequency while temporal dimension is handled loosely on a case to case basis. In proposed proximity filtering method, users' consumption bundles are key drivers. It does not need users' static profile properties rather a latent thematic representation provides user profile. In this work, we close the loop in other two dimensions by attribute similarity learning using temporal proximity filtering. It makes presented system a distinguishing hybrid method due to its naturally induced similarity and dynamic user profiling.}
\label{fig:analogy}
\end{figure}

Dynamic needs of users are largely abstract in nature and can better be learned from their own items consumption history. Current collaborative models fail to address users' inherent needs and liking \cite{wired-netflix} because they ignore either changing user preferences or latent representations. We illustrate an analogy of the presented method with collaborative filtering methods in figure \ref{fig:analogy}. 
Collaborative filtering methods rely on similarity in user profiles and items frequency. User profile properties are largely assumed to be predefined and fixed. Similarities are induced by users and items data matrices \cite{sarwar2001item}. Although these models are popular for predicting next recommended items, they do not have a reliable provision to include temporal needs in their users or attributes feature vectors. They try to handle it by rolling the model temporally but over the same set of similarity inducing static features. Therefore, such methods loose track of temporal dynamics in users' liking or taste. In proposed proximity filtering method, users' consumption bundles are key drivers. We propose that items similarity can be learned from user tastes in a thematic space. It does not need users' static profile properties rather items consumption in a latent thematic space is an embodiment of user profiles. The induced similarity and dynamic user profiling makes the presented model a dynamic hybrid method. 

Traditional recommender methods are not sufficient for diverse and intricate domains like music. Recommending music using fixed qualitative categories like genre, artists etc does not go far and we need to augment the systems with more fundamental compositional properties of music attributes. Moreover, users get attached to particular kinds of music because they start liking the inherent compositional attributes like beats rather than hard-classified genres or artist names. Users listen to music that matches their taste and it might be composed by different artists across genres or same artist or a mixture. A key aspect, we ought to consider, lies in learning the intrinsic structures that make people like a particular song or pair of songs in tandem. It opens up a question - can we handcraft such features which are driven by the dynamic nature of preferences and taste? Songs in a genre might have intrinsic similarities, however, a user might listen to one song from a particular genre but he might not listen to another song from the same genre. It clearly indicates that instead of handcrafting such feature categories, we would be better off learning them implicitly from users' content consumption. It can be a starting point to learn similarities in items' attributes. Such a learning ability provides us a lever to filter recommendable and novel items matching user taste, even if they are newly released.

We fill this gap by inducing similarity between pairs of songs in attribute space. The attributes like beats, loudness, pitch etc have a sequence that characterize songs or other media items. We propose that the items consumed in temporal proximity have underlying structural similarities at a latent level. We describe a measure to  induce similarity by proximity in consumption.
\begin{figure}[h!]
\centering
\includegraphics[width=1.0\linewidth,height=4.05cm]{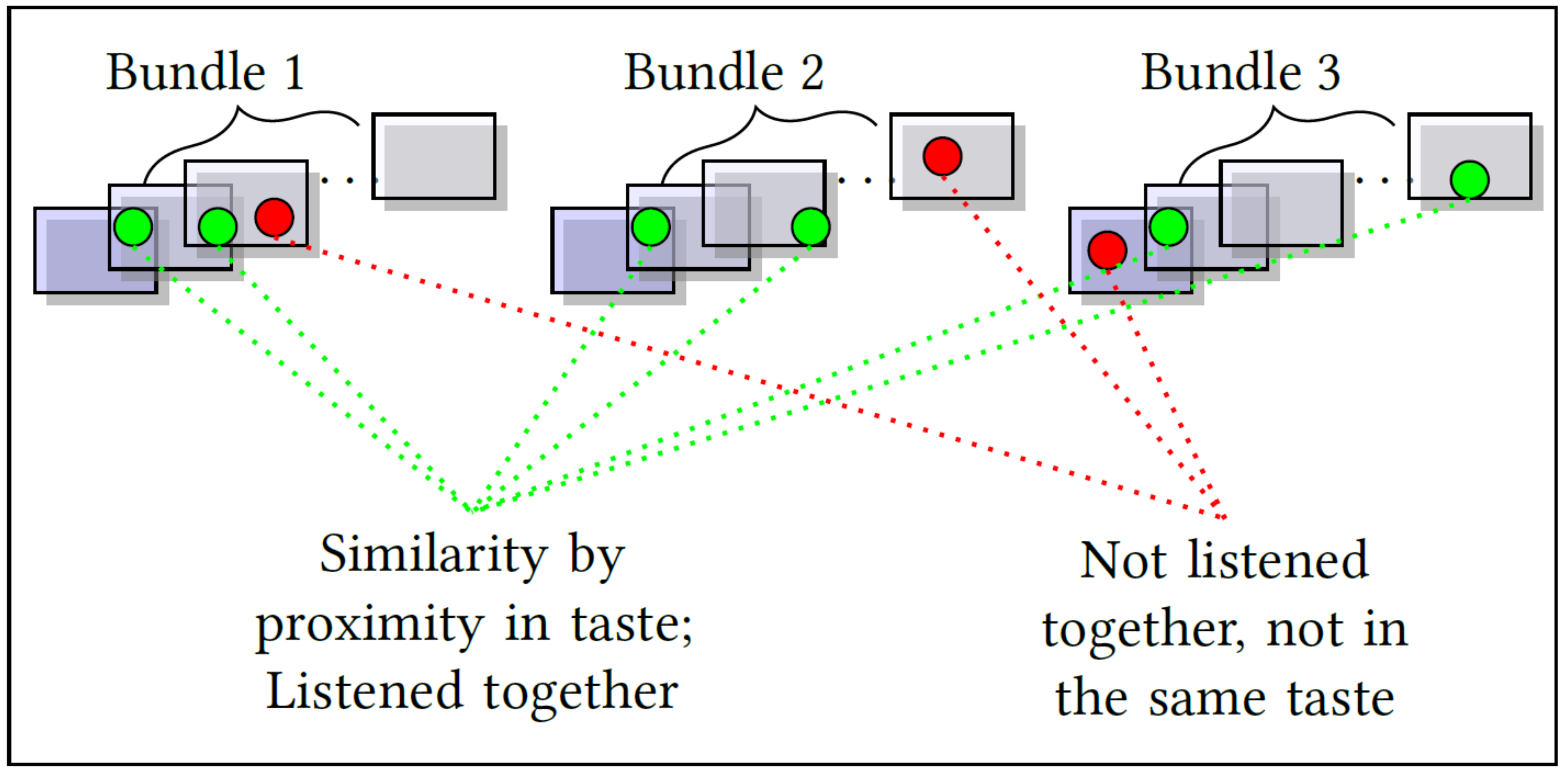}
\caption{Users listen to their favourite songs together, according to their mood and taste. We propose a similarity metric in taste space. Songs listened within a taste proximity are deemed to be more similar than others.}
\label{fig:listening}
\end{figure}
Patterns of items consumption, depending on user's psychological states or liking at a point of time can be represented as a latent theme. Depending on the context like mood, location, or even prevailing political environment of the day, a user might chose different mixture of songs. So, a latent taste theme implicitly reflects consumed items in proximity bundles. Similar to \cite{kumar2017novelty}, we present latent taste themes as analogous to latent topics in Latent Dirichlet Allocation (LDA)~\cite{blei2003latent}. Users' consumption bundles are structures over these latent tastes. This intuition motivates us to present a taste proximity similarity measure. 
By taking an analogy with documents over which topic models are inferred, we take each users' 
listening history as a document, which is a mixture of latent tastes. In this analogy, the individual songs are  words in the text.  By applying LDA over the corpora of user histories (documents), we can get distribution of each user over latent themes. Additionally, we can also infer the probability distribution of each theme (topic) for every song (word) in our history. 

Let the probability distribution of themes for a song $S_x$ be $P(S_x)$ and a song $S_y$ be $P(S_y)$. Then, we can calculate the similarity (cosine similarity) between songs using their theme (topics) distribution given by  
\begin{equation}
{Sim}_t(S_x,S_y) = \frac{P(S_x) \cdot P(S_y)}{ |P(S_x)|  |P(S_y)|}
\end{equation}
Since the songs are composition of a variety of attributes, we can use the taste proximity similarity measure to learn structural similarities in the underlying attributes of paired songs. Users listen to songs in taste bundles as shown in figure \ref{fig:listening}. The taste bundles are sequence of songs listened by a user in streak. For example, a user might want to listen to classicals at the start of the week. If the song pairs listened by users come from the same taste, they are more likely to have some structural similarity than the songs from different tastes. We present this similarity in latent tastes as a learn-able metric in an attribute space. The latent themes model can be run for any parametrized duration, therefore, we call such a representation of proximity in a latent thematic space as \textit{Temporal Proximity Filtering}. 

\emph{Essentially}, users' item consumption in a latent thematic space provides insights into the basic composition of the content itself. We propose that such similarity measure provides a novel opportunity to learn attribute structures.
\section{Attributes Similarity}
\label{sec:attrsimilarity}
Discovery of novel items that match users' tastes among millions of available items is a mammoth task. Filtering from the potential items is usually done either through explicit methods or tracking implicit user behaviors. We propose that proximity filtering in latent thematic space could be a useful way to retrieve interesting and novel potential items that match users' tastes. As demonstrated by Knees et al.\cite{knees2016introduction}, there are two main components in relevant music information retrieval: features and similarity measure.
%
\begin{figure}[ht!]
\centering
\includegraphics[width=1.0\linewidth,height=5.05cm]{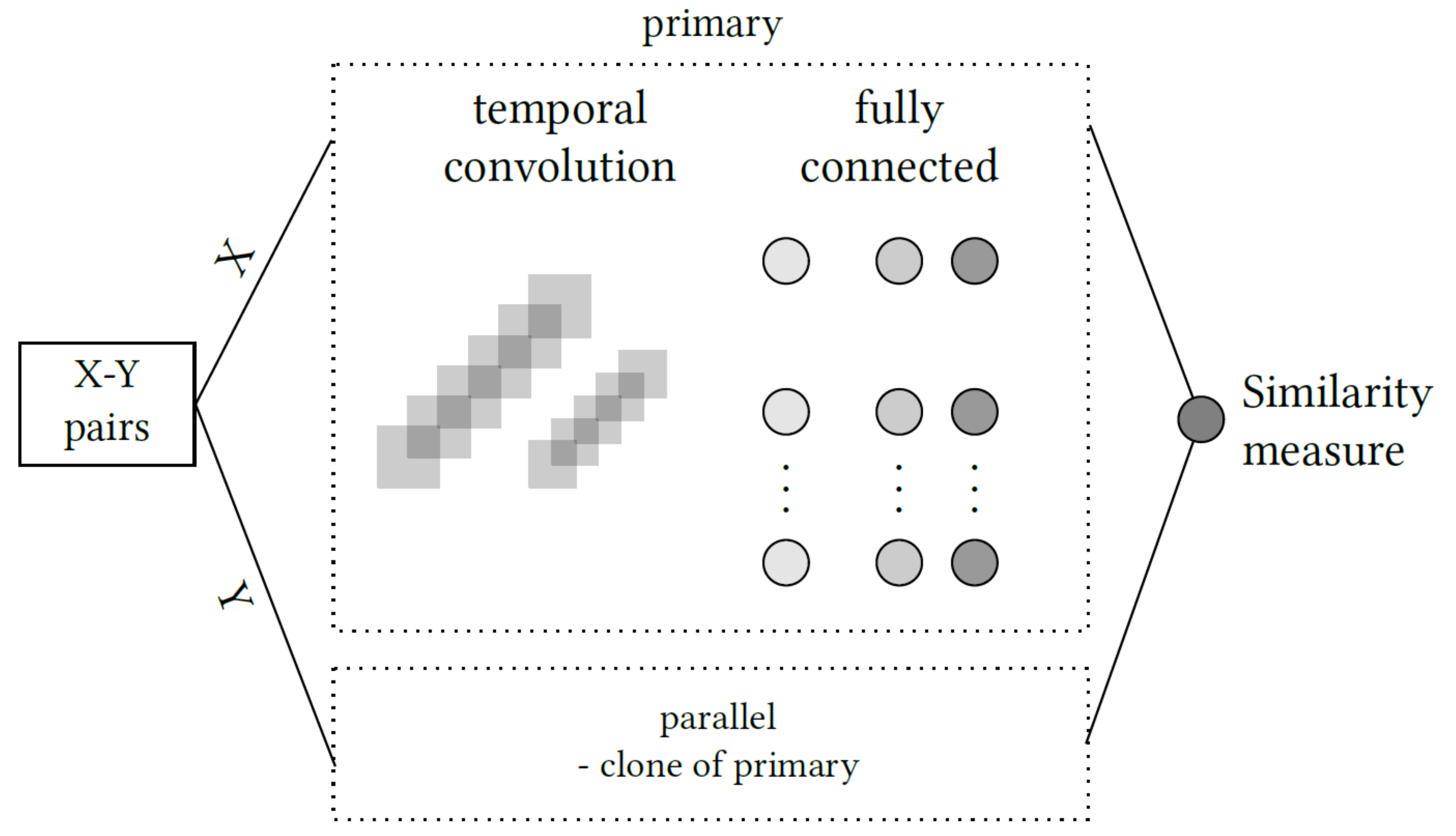}
\caption{ A simple schematic diagram of pairwise attribute similarity metric learning: Song attributes are convoluted for abstract representation and fed to fully connected layers. Cloned network similarly processes second song of the pair and pairwise similarity induced by users' consumption bundles is used as a training signal.} 
\label{fig:attrsim}
\end{figure}
For media content like songs, we need to learn higher level mixture of attributes to understand underlying characterizing structures. We use a convolution network \cite{lecun1995convolutional}\cite{krizhevsky2012imagenet} to extract abstract representations from attributes. Figure \ref{fig:attrsim} illustrates schematic view of learning the attribute similarity for paired items. 

Let X be the input and the number of filters at each level determined by $[m,n]$ kernel configuration, 
$K_{i}$ be a filter with kernel specification at a layer $i$. The output of convolution at a layer $i$, $Cx_{i}$ is given by 
\begin{equation}
\begin{split}
Cx_{i} = f(b_{i} + K_{i} \ast tmP(X_{i}))
\end{split}
\end{equation}
where f is activation function, $b_{i}$ is bias for layer i, tmP is temporal max pooling and $\ast$ represents a convolution operator. We use last $Cx$ to feed a fully connected network given by 
\begin{equation}
\begin{split}
Ox_{l} = f(b_{l} + W_{l}^T Cx)
\end{split}
\end{equation}
where $W_{l}$ represents weight of hidden nodes and $b_{l}$ hidden bias for layer $l$. Similarly, the processing of second item, Y, of the pair is given by 
\begin{equation}
\begin{split}
Cy_{i} = f(b_{i} + K_{i} \ast tmP(Y_{i})) \\
Oy_{l} = f(b_{l} + W_{l}^T Cy)
\end{split}
\end{equation}
with shared parameters between parallel networks. 
We compute similarity between two paired items as 
\begin{equation}
\begin{split}
{Sim}(S_x,S_y) = \frac{{Ox}.{Oy}} { ||{Ox}||||{Oy}||}
\end{split}
\end{equation}
where there are $n$ output units in each of the parallel network, ${Ox}$ and ${Oy}$ are forward pass outputs for the respective item in the pair.

Using proximity filtering, we compute similarities between item pairs based on their consumption in a latent thematic space. Therefore, we use this cosine similarity measure as a teaching signal to train the learning model. We use mean squared error between attribute similarity and thematic teaching similarity as a loss function,  
\begin{equation}
L(S_x,S_y) = \frac{1}{n} \sum_{(x,y) \in S} ||{Sim}(S_x,S_y) - {Sim}_{t}(S_x,S_y)||^2
\end{equation}
where $S$ is the set of pair of songs from items (songs) basket. 
Following this procedure, we learn similarity between paired items and the model trained using an induced similarity is capable of predicting similarity of unseen item pairs. Therefore, given an item, it is feasible to find novel items with similarity in tastes in an attribute space. 

\section{Experiment}
\label{sec:experiment}
In this section, we describe our dataset and the methodology, structure of our convolutional neural network used to co-learn the similarity between the songs in attribute and consumption space.

\subsection{Dataset and pre-processing}
We use the user data from the \last for the temporal music consumption history of 992 users collected by
Celma~\cite{Celma:Springer2010}. The dataset contains complete temporal history of songs listened by the
users from May, 2007 to May, 2009. However, the dataset does not contain the attributes of the songs. 
Musical attributes of songs are provided by 300 GB big Million Song Dataset~\cite{Bertin-Mahieux2011}'s
1 million popular songs extracted by \echo's song analysis~\cite{jehan2005creating}.
These 28 features depict the audio analysis of the songs done using the ``Analyze" tool~\cite{echoapi}
on raw audio to extract these derived features since raw audio data is not publicly accessible to 
researchers because of copyright issues. We only used the 28 audio features like beats, segments, bars, and 
loudness, from the song features listed at \last~\cite{fields}. While these are 28 features, some of them like segments are 2000 dimensional features. With this background, we got around 17000 dimensional space for each song, explained in next subsection.

Since, the MSD songs (\echo) and the
songs from the \last's user data don't have a unique shared identifier, we used artist identifier field   
(missing in some case) in combination with the song title string similarity for correspondence between the 
two datasets. The \last's user data has 1.5 million unique songs, including ads/independent artist 
renderings. Our matching yielded correspondence between 341,039 songs across the two datasets. Out of the rest ~1.16
million songs, ~90\% songs were listened less than once by any of the users. We use these 341,039 unique 
songs for our analysis of the user history, ignoring the rest, since there is no audio analysis data 
associated with them along with low listening frequencies.

\subsection{Procedure}
The proposed procedure, TASTESIM, is illustrated in algorithm \ref{fig:procedure}. We use users' consumption history dataset and song attributes from million song dataset. First, we discover latent themes using an off-the-shelf latent Dirichlet allocation package. We kept the number of themes to 20 and computed user-theme and theme-song distributions. In this experiment, we consider weekly lists that is analogous to weekly documents - imagine a large book divided into sections to be read on a weekly basis or weekly playlists.

\begin{algorithm}[h!]
\begin{algorithmic}[h!]
\caption{$TASTESIM$: Attribute Similarity Learning by proximity in consumption bundles}\label{AttributeLearn}
\\ \textit{Similarity in Taste Space}
\State $\mathcal{W}$ $\gets$ $\{S_1,S_2,S_3 \cdots S_n\}$ set of songs
\State $\mathcal{H}$ $\gets$ Users' consumption history
\State $Puh, Phs$ $\gets$ LDA($\mathcal{H}$,$\mathcal{W}$)
\State $S$ $\gets$ $\{(S_x,S_y) \text{ where } S_x,S_y \in \mathcal{W}\}$
\For {$(S_x,S_y) \in S$}
\State compute ${Sim}_t (S_x,S_y)$ using $Puh,Phs$ (refer Sec. \ref{sec:tempproximity})
\EndFor
\\ \textit {learn attribute similarity}
\State load the network model $\Theta$
\While {not converged or not max-iter}
\State compute ${Sim}(S_x,S_y)$ via parallel + conv + fc layers
\State evaluate loss $L(S_x,S_y)$ (refer Sec. \ref{sec:attrsimilarity}) 
\EndWhile
\label{fig:procedure}
\end{algorithmic}
\end{algorithm}

The dataset provides attributes like bars attributes, segments attributes, pitches, tatums, danceability, duration, and tempo. We consider pairs of songs and compute their cosine similarity scores using distributions from latent theme model. By limiting the scope of the taste model to weekly documents, we ensure that proximity in taste model is analogous to weekly lists. Out of all possible pairs, we used a subset of 10K pairs for taste induced similarity. We use temporal convolution to abstract features, weights act as kernel filters and are learned during network training. Filter size is kept as 5. Following \cite{scherer2010evaluation}, we use temporal max pooling and ReLU as activation in intermediate hidden layers \cite{dahl2013improving}. The abstracted attributes are fed to fully connected layers with 800 and 600 hidden units. The number of output units are kept as 20 in correspondence with the number of themes. Thus, we show that taste similarity in thematic space can be used to induce similarity in learning item properties matching users' tastes.

\section{Results and Discussion}
\label{sec:discussion}
We discuss the existence of temporal structure in consumption bundles and taste induced similarity for a publicly available users' song consumption dataset and million songs dataset. 

\subsection{Temporal Consumption Pattern}
\begin{figure}[h!]
\centering
\includegraphics[width=0.9\linewidth,height=4.65cm]{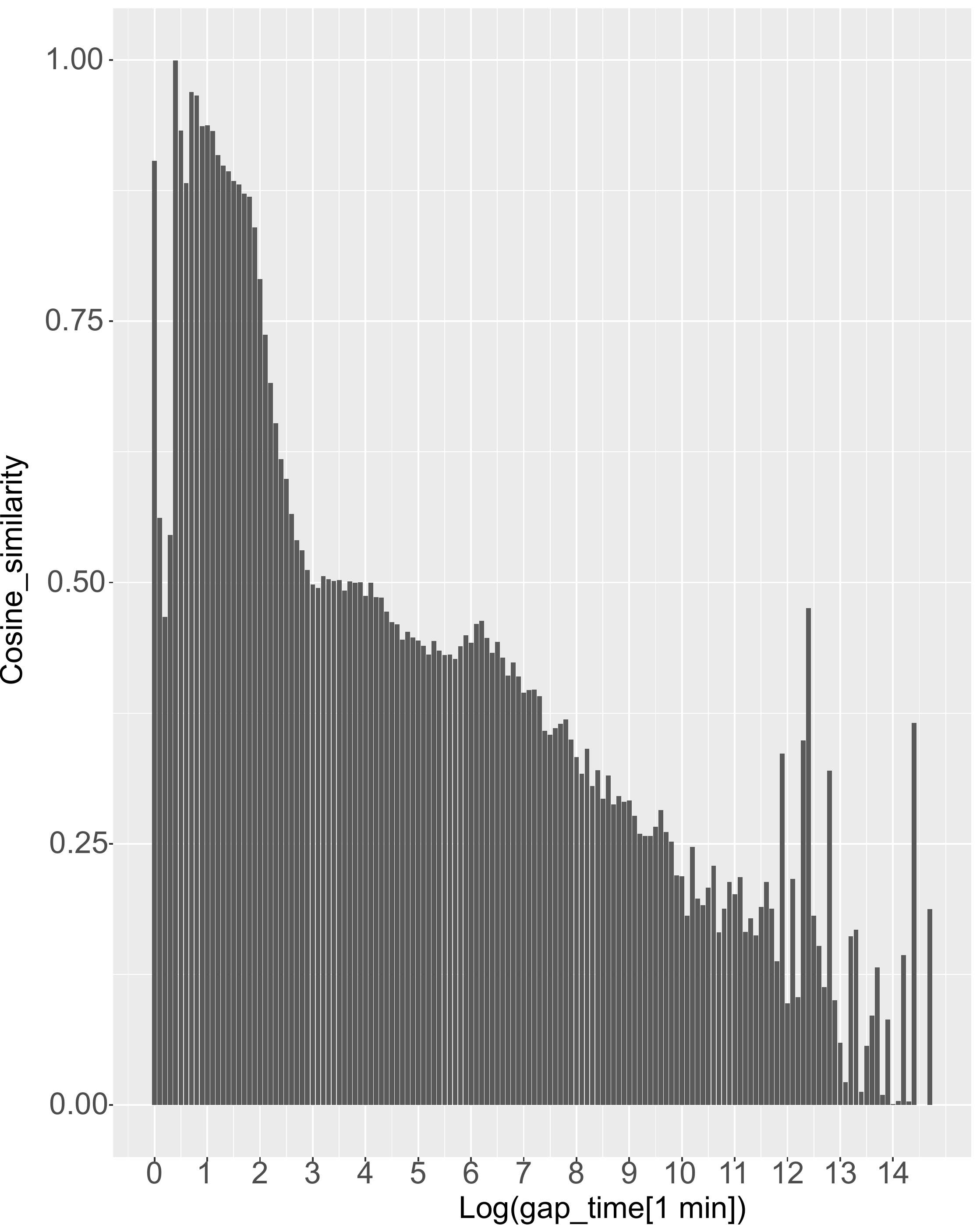}
\caption{Cosine similarity across $log(\Delta T)$ with one minute time scale. Higher similarity scores for lower gap times demonstrate that proximal consumption of similarity in taste does exist.}
\vspace*{-3mm}
\label{fig:cosineSimilarity}
\end{figure}

We first examine if temporal structure does exist in user consumption patterns in taste space. We compute gap time between two songs and their similarity score as given in sec \ref{sec:tempproximity}. This score indicates how similar are two songs in taste space. We plotted gap time with similarity scores and from figure \ref{fig:cosineSimilarity}, it is evident that similarity scores are higher for items consumed within a gap of couple of hours. A higher score in taste means that those pairs of songs have more commonalities in consumption and lower gap time means that they are listened more in tandem. 

\begin{figure}[h!]
\centering
\includegraphics[width=0.9\linewidth,height=4.65cm]{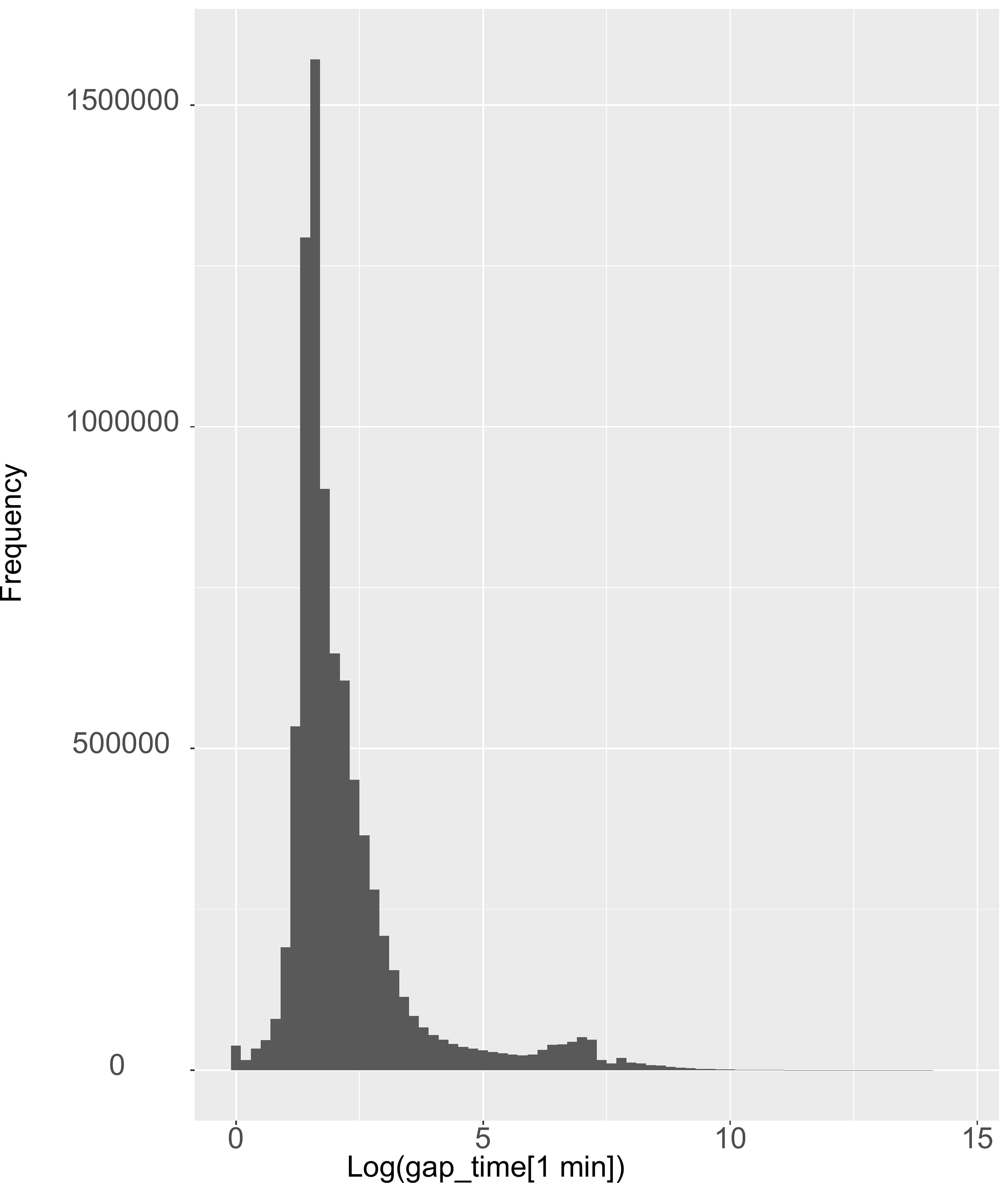}
\caption{Frequency of $log(\Delta T)$ with one minute time scale.}
\vspace*{-3mm}
\label{fig:frequency1min}
\end{figure}

Moreover, as illustrated in figure \ref{fig:frequency1min}, such pairs are not only listened in temporal proximity but also listened more frequently. A comparatively lower frequency on the left side of the plot in figure \ref{fig:frequency1min} and gradual decrease in the frequencies on right imply that people might not listen to pairs in immediate continuity but in a short period, it is more likely that they return to their favourite pairs of songs. It indicates that temporal proximity structure does exist and provides us an important insight into user tastes. 

\begin{figure}[h!]
\centering
\includegraphics[width=0.9\linewidth,height=4.65cm]{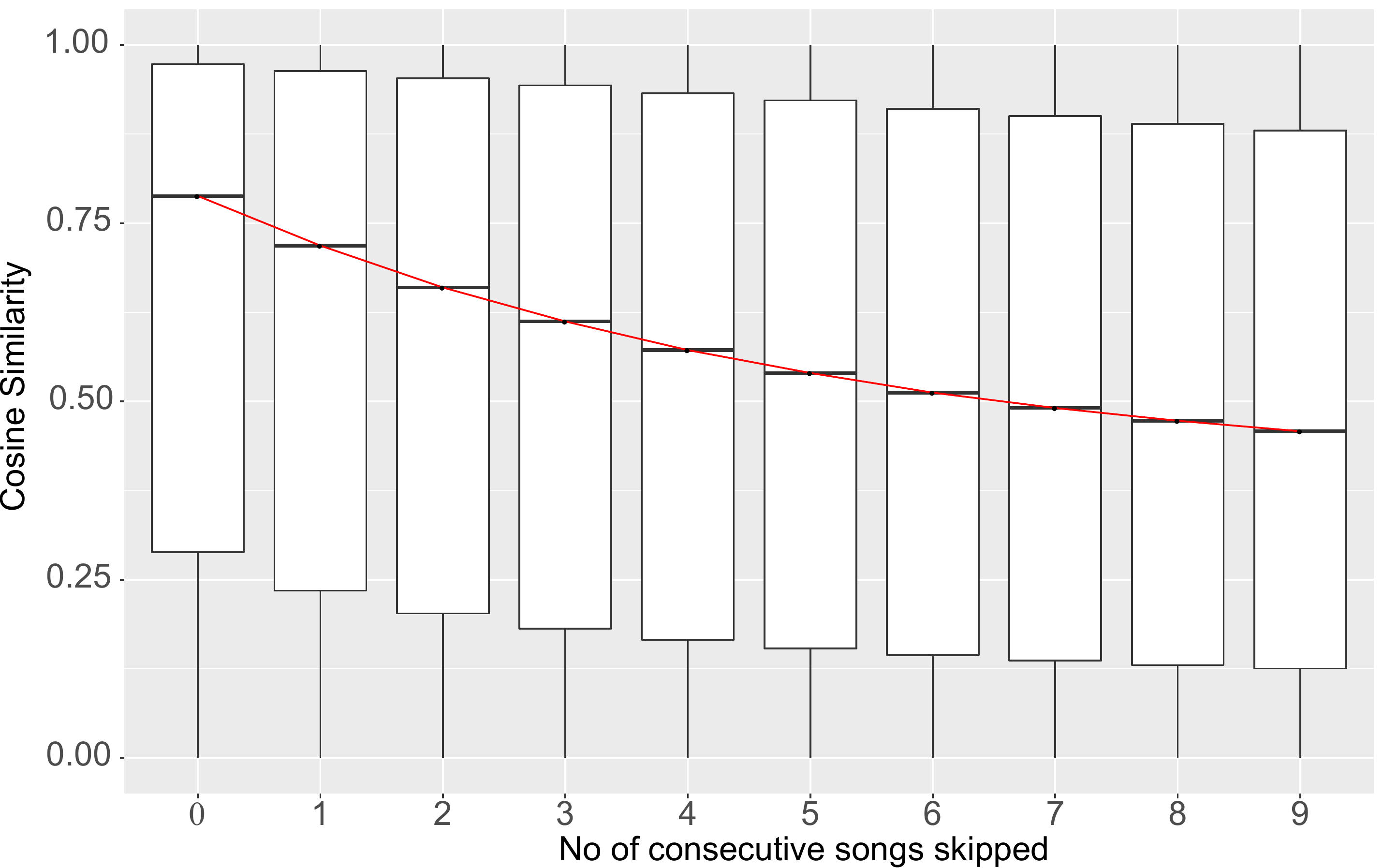}
\caption{Boxplot showing cosine similarity distribution of song taste listened by users by skipping {\it n} consecutive songs given on the x axis. Horizontal axis represents the number of consecutive songs skipped for calculating the cosine similarity between the songs for a user. Vertical axis represents the cosine similarity values. }
\label{fig:nskip}
\end{figure}

Traditional methods use external user attributes which might not be a good indicator of user tastes and preferences. We have illustrated that user tastes in a period are manifested in grouping of songs listened or consumed in temporal proximity. A similar argument can be made for other media consumption domains. Existence of such proximal structure can be leveraged to induce valuable knowledge in discovering novel items that a user might prefer. 
We now examine skip level structures, in other words, existence of Markovian structure in music listening in more details. We construct pairs of songs using n-skip levels where {\it n} indicates number of consecutive songs skipped over. For example, 0-skip would mean that we consider every consecutive song, 1-skip would mean that we consider every alternate song. A similarity score using taste model is computed for each of the n-skip pair. Boxplot (mean values with one-Standard-deviation) in figure \ref{fig:nskip} illustrates cosine similarity distribution of song taste listened by users by skipping {\it n} consecutive songs. From this observation, we can say that temporal proximity structures exist and we can say that users listen to their favourite songs in temporal proximity according to their tastes.

This is further evidence that users listen to songs matching their tastes in temporal proximity provides a novel way to induce similarity. Since similarity between pair of songs is derived from users' content consumption, it is viable to think that there should be an underlying implicit structure in attributes space. For example, if two items frequently co-occur within the same theme, they are more likely to have some underlying properties that encourage their co-consumption. Therefore, we should naturally exploit this implicit knowledge derived from users' tastes to find items. 
\subsection{Attribute Similarity Learning through Proximity Similarity }
\begin{figure}[h!]
\centering
\includegraphics[width=0.9\linewidth,height=4.65cm]{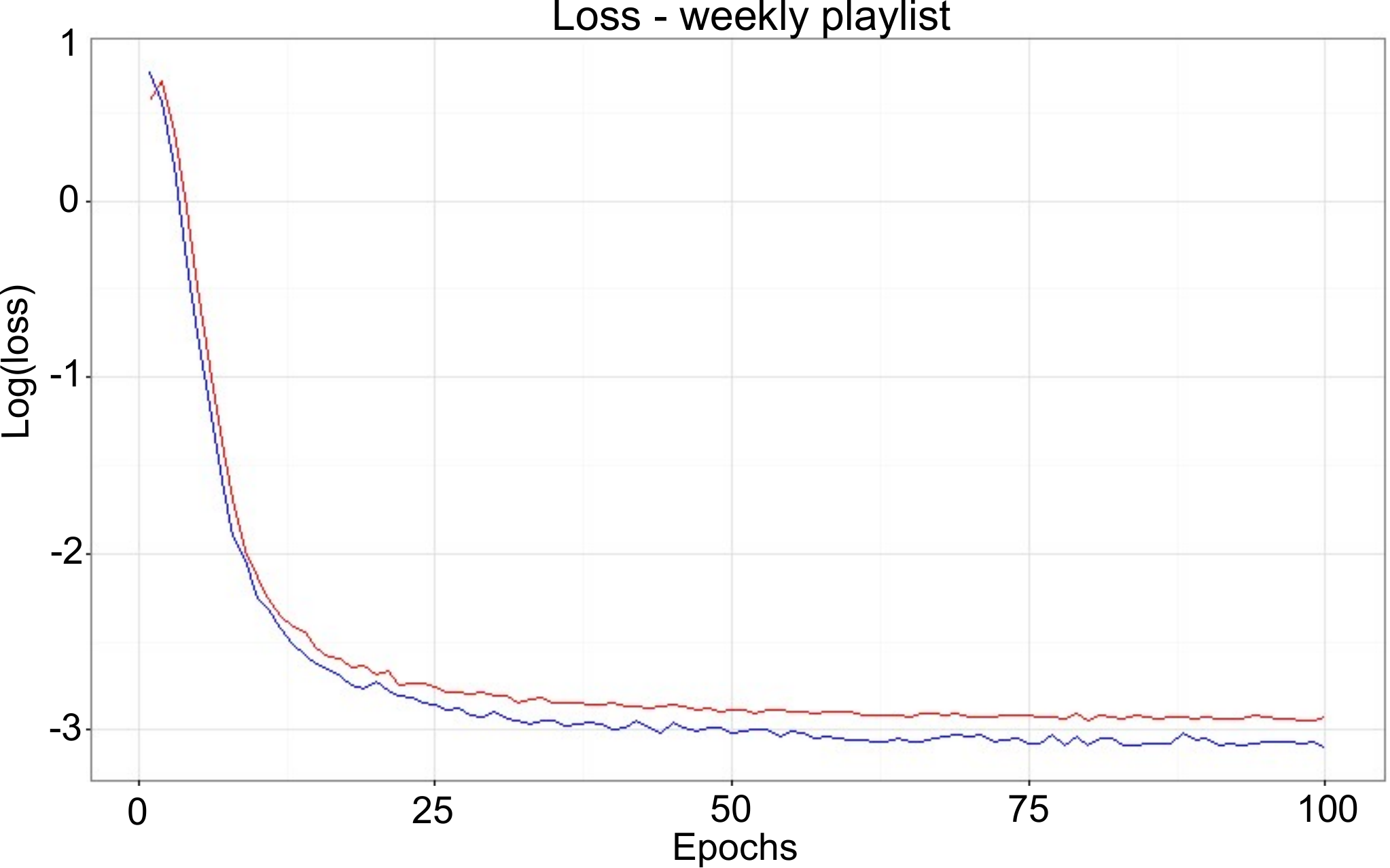}
\caption{Illustration that proximity driven similarity can be used to induce similarity in item-pairs in attribute space. Red: train and blue: validation loss}
\label{fig:learnSimilarity}
\vspace*{-3mm}
\end{figure}

Taking a cue from play lists, we consider temporal proximity in bundles. We create weekly listening patterns of the users from one year consumption data, \textit{i.e.}, 52 maximum possible weekly play-lists. For each user we only consider a weekly play-list for which they listened to a song. The intuition behind using the weekly play-lists is that the music streaming platforms usually publishes weekly charts that are used by the users to refer to the latest trendy songs. In other words, we infer users' weekly music taste from this play-list (document) using LDA. Training the network model, fig \ref{fig:learnSimilarity}, using taste or thematic similarity shows that induced taste similarity can be used to learn similarities between item-pairs in attribute space. We get a MSE of the order of $10^{-3}$ on the test set, which reflects the difference in similarity between a pair of songs from the users' taste space and the learnt similarity calculated by the deep network from the song attributes. Hence, our deep network is able to learn a latent space of item-item similarity in the attribute space that mimics the dynamics of user driven item-item similarity.

In a nutshell, we showed that users consume media content in temporal proximity of consumption bundles which reflect users tastes. Instead of using indirect and non-reliable methods on trying to assess what a user might prefer, we propose that using these consumption bundles is a viable option to induce knowledge of users taste in attribute space. This novel view could be beneficial in developing users tastes and preferences driven recommender systems. 

\section{Conclusion}
\label{sec:conclusion}
Identifying novel items is a persisting challenge in recommendation systems. While a lot of strides have been made in improving the quality of recommended items, filtering recommendable items that match user tastes and dynamic preferences is an area needing more attention. Traditional methods suffer from ``long tail" effects leading to suppression of items very similar to popular ones, but not listened often. In this work, we attempt to address this issue by presenting a novel method by inducing users' taste driven temporal proximity similarity measure into attribute space without user meta data or any explicit feedback. 

We demonstrated that temporal structure exists in users consumption behavior and users consume their favourite content, matching latent theme or taste, in temporal proximity of consumption bundles. We presented an induced similarity metric from latent theme (or taste) model and showed that it is possible to harness users' taste induced similarity for item-pairs in attribute space. 

Therefore, temporal proximity in latent thematic or taste space provides a fundamentally novel view in unravelling what users implicitly prefer instead of relying on outside knowledge. This view has potential in devising recommendation systems by finding novel items that match user taste in attribute space. 

\bibliographystyle{ACM-Reference-Format}
\balance
\bibliography{music}
\end{document}